# Diamine Surface Passivation and Post-Annealing Enhance Performance of Silicon-Perovskite Tandem Solar Cells

*Margherita Taddei,[1] Hannah Contreras,[1] Hai-Nam Doan,[2,3] Declan P. McCarthy,[4] Seongrok Seo,[2] Robert J. E. Westbrook,[1] Daniel J. Graham,[5] Kunal Datta,[6] Perrine Carroy,[7] Delfina Muñoz,[7] Juan-Pablo Correa-Baena,[6] Stephen Barlow,[8] Seth R. Marder,[4,9,8] Joel A. Smith,[2] Henry J. Snaith,[2] David S. Ginger\*[1,10]*

1 Department of Chemistry, University of Washington, Seattle, WA 98195, USA

2 Department of Physics, University of Oxford, Oxford, OX1 3PU, U.K.

3 Institute of Materials Science, Technical University of Munich, München, 85748, Germany

4 Department of Chemistry, University of Colorado-Boulder, Boulder, CO 80309, USA

5 Department of Bioengineering, University of Washington, Seattle, Washington 98195

6 School of Materials Science and Engineering, Georgia Institute of Technology, North Ave NW, Atlanta, Georgia 30332, USA

7 Université Grenoble Alpes, CEA, Liten, Campus Ines, 73375 Le Bourget du Lac, France

8 Renewable and Sustainable Energy Institute, University of Colorado-Boulder, Boulder, CO 80309, USA

9 Department of Chemical and Biological Engineering and Materials Science Program, University of Colorado-Boulder, Boulder, CO 80309, USA

10 Physical Sciences Division, Physical and Computational Sciences Directorate, Pacific Northwest National Laboratory, Richland, Washington 99352, United States

\* Corresponding author

*Abstract:*

We show that the use of 1,3-diaminopropane (DAP) as a chemical modifier at the perovskite/electron-transport layer (ETL) interface enhances the power conversion efficiency (PCE) of 1.7 eV bandgap FACs mixed-halide perovskite single-junction cells, primarily by boosting the open-circuit voltage ($V_{OC}$) from 1.06 V to 1.15 V. Adding a post-processing annealing step after $C_{60}$ evaporation, further improves the fill factor (FF) by 20% from the control to the DAP + post-annealing devices. Using hyperspectral photoluminescence microscopy, we demonstrate that annealing helps improve compositional homogeneity at the top and bottom interfaces of the solar cell, which prevents detrimental bandgap pinning in the devices and improves $C_{60}$ adhesion. Using time-of-flight secondary ion mass spectrometry, we show that DAP reacts with formamidinium present near the surface of the perovskite lattice to form a larger molecular cation, 1,4,5,6-tetrahydropyrimidinium (THP) that remains at the interface. Combining the use of DAP and the annealing of $C_{60}$ interface, we fabricate Si-perovskite tandems with PCE of 25.29%, compared to 23.26% for control devices. Our study underscores the critical role of chemical reactivity and thermal post-processing of the $C_{60}$/Lewis-base passivator interface in minimizing device losses and advancing solar-cell performance of wide-bandgap mixed-cation mixed-halide perovskite for tandem application.

*TOC:*

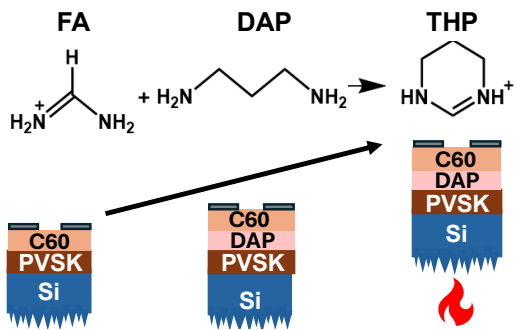

**Introduction**

A tandem solar cell is composed of a narrow band-gap bottom cell and wider band gap top cell.[1,2] This design can surpass the power conversion efficiency (PCE) of a single junction by more efficiently harvesting different regions of the solar spectrum, reducing charge carrier thermalization losses to produce cells with higher voltages. Silicon cells, with a bandgap of ~1.1 eV, can be used as the bottom cell and metal halide perovskites can be suitable active layers for the top cell, with an ideal bandgap of ~1.7 eV for the perovskite.[3–5] The development of Si-perovskite tandem solar cells has gained significant attention, with a record-breaking PCE of 33.9% achieved in 2023,[6] which exceeds the single-junction limit for a Si cell, but is still below that theoretical >43% radiative efficiency limit of an ideal Si-perovskite tandem.[1,6] Challenges persist, particularly regarding the development of perovskite top cells with band gaps above 1.7 eV. The mixed bromide-iodide compositions generally used for 1.7 eV perovskites often exhibit undesirable photoinduced phase segregation,[7,8] leading to lower band-gap domains in which charges are "funneled",[9–11] decreasing the open-circuit voltage ($V_{OC}$) of the solar cell.[12–14] Furthermore, issues arise at the interface between the perovskite absorber and the charge-transporting layers, especially in such wide band-gap perovskite cells. Their deeper valence band maxima (VBM) and shallower conduction band minima (CBM) compared to more typical ~1.6 eV bandgap perovskite compositions result in unfavorable energetic offsets between the perovskite band edges and the transport levels of commonly used electron- (ETL) and hole-transport layers (HTL).[15,16] Fullerenes have been widely used as electron-transport layers (ETLs) in p-i-n perovskite solar cells due to their suitable energy alignment, typically high electron mobility, and ease of processing in vertical device stacks.[17] However, the interface between perovskite and fullerenes is a primary source of efficiency loss in devices.[18–20] This loss primarily stems from charge recombination, particularly the recombination of electrons extracted to the fullerene layer with holes at the perovskite surface.[18,21] The process of charge trapping at the perovskite/ETL interface is also associated with the presence of current-voltage hysteresis in perovskite devices, by which differences are observed in the J-V curves measured in forward and reverse scan directions.[22] Shao and co-workers have shown that adding a layer of the $C_{60}$-derivative [6,6]-phenyl-$C_{61}$-butyric acid methyl ester (PCBM) between the perovskite and $C_{60}$, followed by an annealing step, resulted in hysteresis-free J-V curves.[23] They attributed the lack of hysteresis to PCBM filling grain boundaries of the perovskite layer after annealing, thus lowering the trap density at the perovskite/ETL interface. However, the use of PCBM is less desirable for future Si-perovskite tandem commercialization due to its sensitivity to light and air exposure, as well as stronger parasitic optical absorption and higher cost than for $C_{60}$.[24] Other perovskite/ETL interface passivation strategies have been implemented to mitigate interface recombination based on various Lewis base molecules.[25,26] The deposition of a Lewis base at the perovskite top surface can effectively enhance the photoluminescence quantum yield and reduce surface recombination velocity by passivating surface trap states in the perovskite,[27–29] and can reduce interfacial recombination of holes in the perovskite layer with electrons in the $C_{60}$ layer by electronically decoupling the two layers.[30,31] Diamines have been shown to be a promising class of molecules to passivate bulk and surface defects in various perovskite compositions by being used as additives and surface passivators.[10,32,33] While these passivation strategies have led to overall improvements in solar-cell efficiency and stability, the presence of hysteresis and $V_{OC}$ losses can persist when $C_{60}$ is deposited onto the surface-passivated perovskite layer.[20,34] In this work we investigate the combined effect of using an amine-based surface passivator, 1,3-diaminopropane (DAP), and post-processing annealing of $C_{60}$ and a 1.7 eV mixed-cation mixed-halide perovskite in Si-perovskite tandem devices.

**Results and Discussion**

First, we fabricated single-junction p-i-n perovskite solar cell devices using Me-4PACz as a self-assembled monolayer (SAM) hole-transport layer (HTL) and deposited on a sparse interlayer of alumina nanoparticles to improve the wettability of the perovskite on top of the SAM.[10,35] We use a mixed-cation mixed-halide perovskite composition $FA_{0.83}Cs_{0.17}Pb(I_{0.75}Br_{0.25})_3$ which has a bandgap of ~ 1.70 eV, ideal for Si-perovskite tandems.[10] We then spin coat a solution of 1,3-diaminopropane (DAP) in isopropanol with the optimized concentration of 0.75 mM on top of the perovskite. **Figure S1** shows the device performance from a series of DAP concentrations. After we deposit DAP, we evaporate a layer of $C_{60}$ (thickness of 30 nm) as the electron-transport layer (ETL) and bathocuproine (BCP, thickness of 8 nm) as a hole-blocking layer. We anneal for 1 minute at 150 °C after deposition of the ETL and BCP layers before depositing the silver top electrode.

In **Figure 1a, b, c** and **d** we show the performance results of solar cells (active area 0.0453 cm$^2$) made without the DAP deposition step (denoted here after as "Control"). The performance is improved when introducing the DAP surface passivator in the stack ("DAP") and annealing after $C_{60}$ evaporation (denoted here after as "Control + anneal" and "DAP+ anneal"). The full device architectures are shown in **Figure S2**. The maximum PCE recorded for the Control device was 15.49% under forward scan which improved to 18.24% with DAP and further to 19.11% after DAP + annealing. The fill factor improved the most, increasing from 63.4% in the control and 73.3% with DAP, to values over 75% after annealing. In **Figure S3** we show that the annealing step also has a positive impact when using other amines, such as lysine and polyallylamine, as interlayers. However, we achieved the best results using DAP as the interlayer for this perovskite composition. The J-V curves of a typical device for each processing condition are shown in **Figure S4** and the mean and standard deviation values of devices of Figure 1 listed in **Table S1**. We monitored the stability of the control and DAP-treated devices by measuring the performance of encapsulated devices over a period of 324 hours in 1 Sun-equivalent constant illumination under open-circuit conditions at 85 °C (ISOS L-2).[36] In **Figure S5,** we show that the PCE, $V_{OC}$ and FF are better retained in the control and DAP-treated devices that experienced the annealing step. The $J_{SC}$ did not change. The stability of the $V_{OC}$ and steady improvement of the FF after aging suggests that the annealing is a promising step for improving device stability as well as performance. Moreover, we investigated whether the device performance increased with annealing prior to the evaporation of $C_{60}$. **Figure S6** shows the performance of control and DAP-passivated devices that were either annealed before ("ANNEAL PVK") or after the evaporation of C60 and BCP ("ANNEAL C60"). The highest power conversion efficiency (PCE) is achieved with DAP and annealing after C60 evaporation, demonstrating that these two processes are the most effective combination in our study.

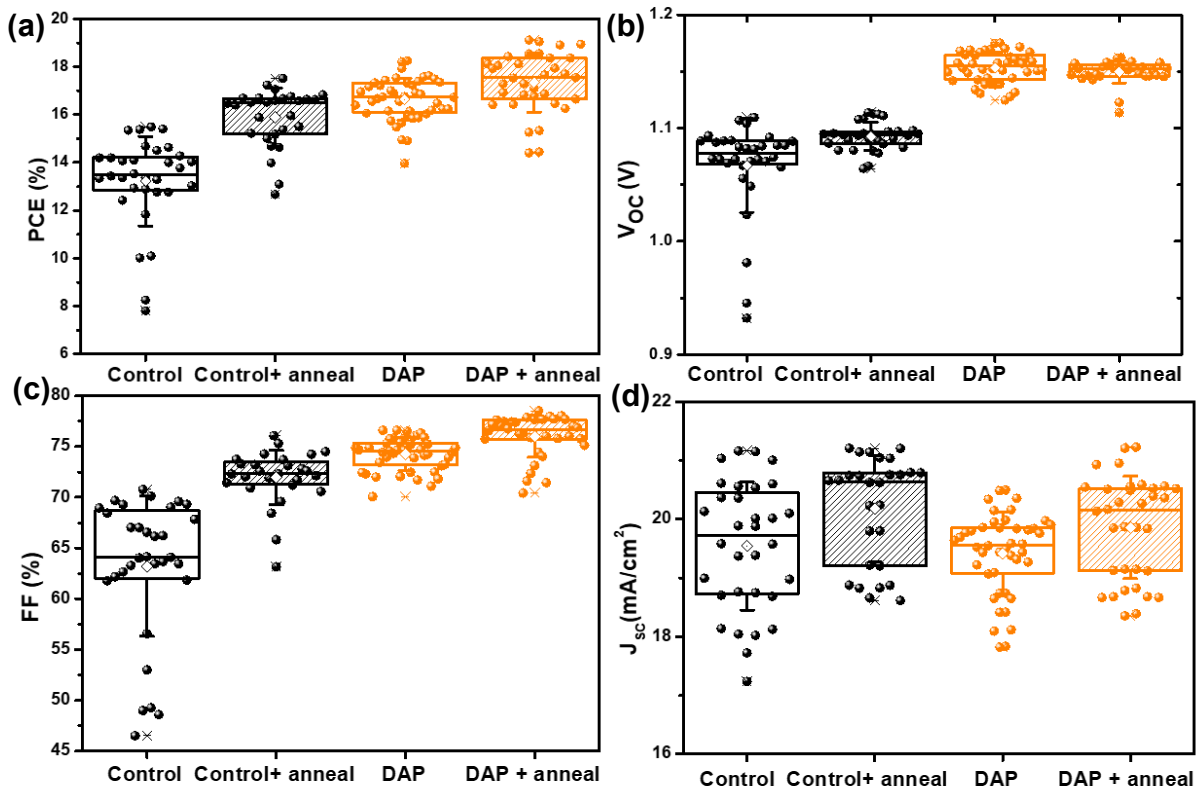

*Figure 1*: a) Power conversion efficiency (PCE %) values, c) open-circuit voltage ($V_{OC}$), d) fill factor (FF) and e) short-circuit current density (Jsc) for the control device ("Control", black), DAP-treated ("DAP", orange) and Control and DAP-treated with a further annealing step at 150 °C for 1 minute after $C_{60}$ deposition ("Control + anneal", black, striped and "DAP + anneal", orange, striped).

To better understand the impact on the FF, in **Figure S7** and **S8** we examine $C_{60}$ coverage by recording photoluminescence (PL) maps of the devices. We see that the control device has more "bright spots" with higher PL and longer lifetime, which we attribute to the absence of $C_{60}$, and associated PL quenching (**Figure S9**). The devices treated with DAP and annealing have fewer bright spots, indicative of better $C_{60}$ coverage. The fewer bright spots in DAP-treated and annealed samples suggests that the use of DAP and subsequent annealing of the $C_{60}$ interface may reduce the number of pinholes in the ETL layers resulting in higher FF. To gain insights into how the

DAP deposition step and the annealing process affect the optoelectronic properties, we conducted hyperspectral PL microscopy on the full devices (**Figure 2**). By using a hyperspectral microscope, we obtain a PL spectrum at every pixel with high spatial resolution. This technique has been used to spatially resolve heterogeneity in perovskites and other semiconductors.[10,37–40] Here, we capture hyperspectral images after excitation from the top surface of the perovskite, comparing the different treatments. We use a mercury halide lamp in the 350 – 450 nm range, which has penetration depth of ~50 nm, at 130 mW/cm$^2$ excitation intensity for ~5 minutes.[41]

We include the PL intensity maps in **Figure S10** and the average PL spectra in **Figure S11**, which show higher PL intensity for the DAP-treated full stack devices analyzed from the top interface compared with the untreated devices. In **Figure 2** we show the PL peak emission wavelength of the samples. The presence of "wrinkled" regions of redshifted and blueshifted PL has been previously shown for these wide gap mixed cation mixed halide compositons.[7,9,10,42,43] In **Figure S12 a-d** we show the PL peak wavelength map excited and collected from the top side Glass/BCP/C$_{60}$/perovskite interface for the control, control + anneal, DAP and DAP + anneal devices. The images show similar values of PL peak wavelength with values of 734 nm for the control, and control + anneal, 733 nm for DAP and 732 nm for DAP + anneal device. The distributions of the PL peak wavelengths are shown as histograms in **Figure S12 e**. In **Figure 2 a-d** we show the PL peak wavelengths map of the Glass/ITO/SAM/perovskite interface of the same four devices. For both the control and DAP-treated devices the PL peak taken from the Glass/ITO/SAM interface is slightly more redshifted compared to the top with PL peak wavelengths of 742 nm for the control and 737 nm for the DAP-treated one. This slight redshift suggests that for both the control and DAP-treated device the perovskite layer is vertically heterogeneous at the two interfaces, being more iodide-rich at the bottom Glass/ITO/SAM contact and more well intermixed at the top interface with $C_{60}$. After annealing the back interface PL peak wavelength blueshifts, with an average PL peak at 736 nm for the control + anneal and 733 nm in the DAP + anneal (**Figure 2e**). This shift indicated that the perovskite layer after annealing has a more homogeneous wide bandgap composition throughout the device which might explain why the FF is improved only in this case potentially due to better electrical contact and/or energy alignment at both the top and bottom interlayers.

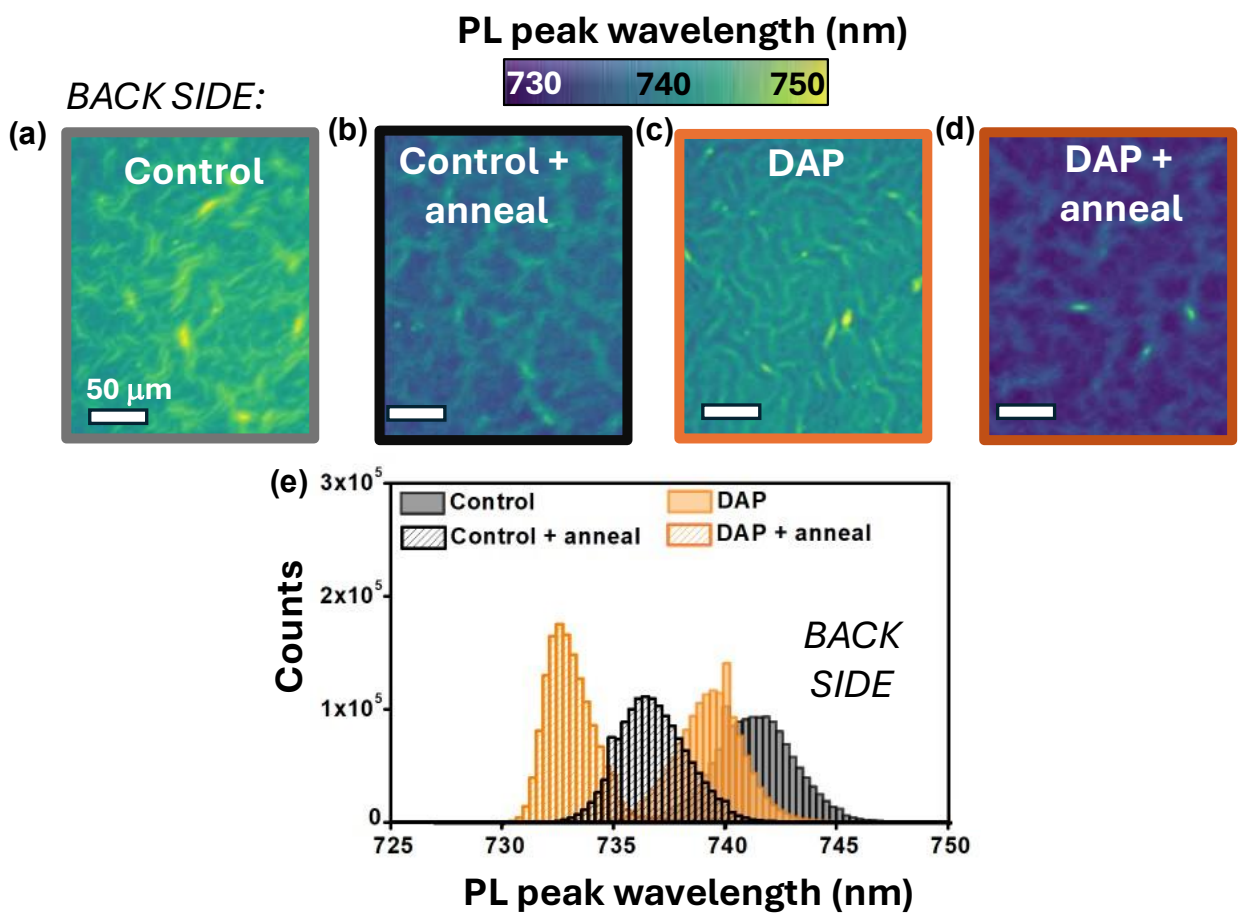

*Figure 2: a) PL peak wavelength map taken from the back ITO/SAM/perovskite interface of the control, b) control + anneal, c) DAP and d) DAP+ anneal solar cells. e) Histograms of the PL peak distribution from a-d.*

To better understand the chemistry behind DAP's effect on the surface we performed time-of-flight secondary ion mass spectrometry (ToF-SIMS). We analyzed a set of three samples: one control sample, and two treated with DAP 0.75 mM and 10 mM solutions. DAP 0.75 mM is the concentration we found to be most effective for improving the efficiency of solar-cell devices, while we use 10 mM to observe possible trends in the formation of new molecular species. We have previously investigated the chemical reactivity of benzylamine (BA) and ethylenediamine (EDA) in solution with $^1$H and $^{13}$C-NMR spectroscopy and in films with ToF-SIMS when used as an additive in the same $FA_{0.83}Cs_{0.17}Pb(I_{0.75}Br_{0.25})_3$ perovskite composition.[10,44] In those studies, we found that BA and EDA will react quickly and quantitatively with $FA^+$ in the perovskite precursor to form larger A-site cations which is subsequently incorporated into the thin film.

In **Figure 3a,** we illustrate a potential mechanism for the reaction of DAP with FA at the film surface, in analogy to what we have observed previously with EDA-treatment.[10] The reaction between DAP and FA would result in the formation of a six-membered ring in which the two nitrogen atoms of a $FA^+$ ion are linked by a propan-1,3-diyl bridge, i.e., 1,4,5,6-tetrahydropyrimidinium ($THP^+$), along with the elimination of two equivalents of volatile ammonia. In **Figure S13**, we describe this possible reaction mechanism in more detail. In **Figure 3b** we show the ToF-SIMS depth profiling results combined with 2D maps of $FA^+$ (m/z=45), $THP^+$ (m/z=85) and Si (m/z=28) mass fragments.[45] We observe an increase in the $THP^+$ signal at higher concentrations. This trend suggests that DAP is converted to $THP^+$ upon contact with $FA^+$ on the surface. Based on Figure 3b, we can see qualitatively that the $THP^+$ signal increases with higher DAP concentrations (10 mM), and it is present on the surface of both films after DAP 0.75 mM and 10 mM surface treatments. After converting the sputter time to film depth (assuming a total film thickness of 500 nm), we see that $THP^+$ is present in the first ~70 nm (**Figure S14**). We analyzed the crystal structure using X-ray diffraction (XRD) shown in **Figure S15**. The XRD patterns of the samples with excess DAP show no additional peaks from secondary phases. We also conducted Grazing Incidence Wide-Angle X-ray Scattering (GIWAXS) measurements to further study the surface and bulk film structures using difference X-ray incidence angles. In **Figure S16** we show the GIWAXS patterns of the control samples and with addition of DAP 0.75, 10 and 75 mM. We see no detectable 2D or polytype structures emerge with the addition of excess DAP.

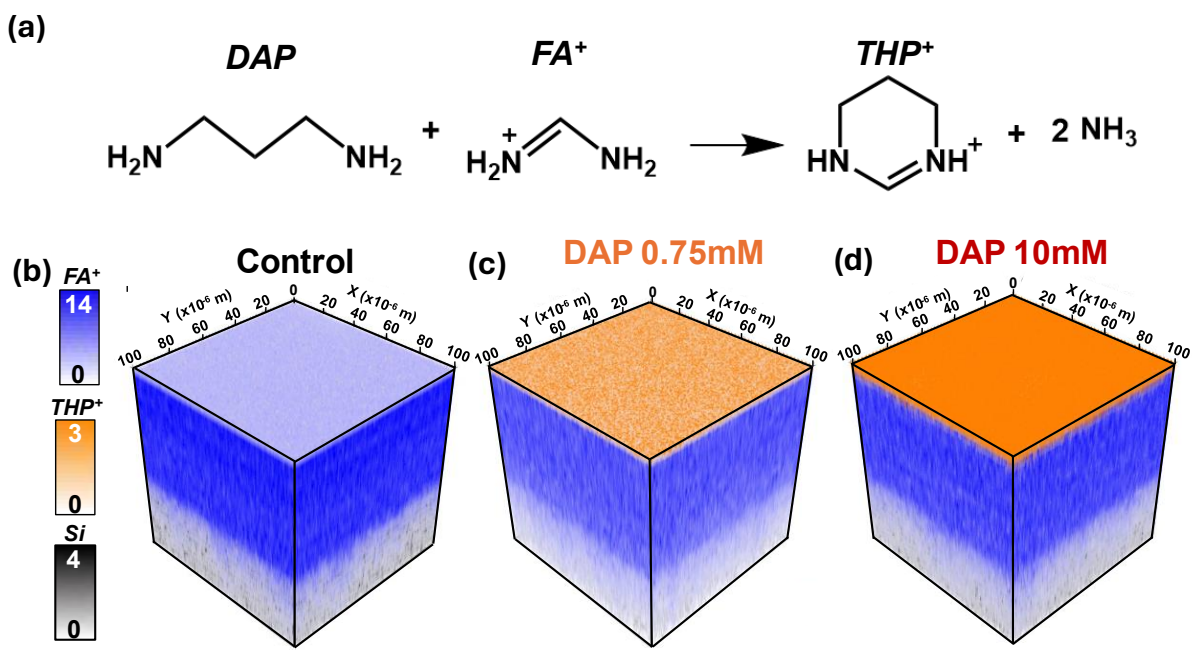

*Figure 3:* a) Reaction scheme of 1,3-diaminopropane (DAP) surface passivator with formamidinium (FA$^+$) cations present in the perovskite lattice to form stoichiometrically 1,4,5,6-tetrahydropyrimidinium (THP$^+$) and ammonia (NH$_3$$^+$). b) ToF-SIMS depth profiling of a control film, and after surface treatment with c) DAP 0.75 mM and d) DAP 10 mM solution in IPA. The molecular fragment of FA (CH$_5$N$_2$, m/z = 45.04) is shown in blue, THP (C$_4$H$_9$N$_2$, m/z = 85.09) in orange and Si (m/z=27.97) in black.

The improved compositional heterogeneity of the DAP + anneal devices suggest that this method would be suitable for achieving higher $V_{OC}$ and FF in Si-perovskite tandem devices where the halide segregation of perovskite compositions of > 1.7 eV can be a limiting factor. We therefore made Si-perovskite tandems with different top layers: the control perovskite layer, DAP-treated perovskite, and perovskite that is DAP-treated and annealed post-processing. **Figure 4a** shows the overall device architecture. **Figure 4b** shows the J-V curve of our control Si-perovskite tandem and in Figure **4c** the one made with DAP passivation and annealing after $C_{60}$ deposition for 1 min at 150 °C. **Figure S17** shows the Si-perovskite tandems JV curves for devices made with DAP only without annealing along with the average performance parameters in **Figure S18** and **Table S2**. We find that the $V_{OC}$ progressively increases for the tandems containing DAP and DAP + anneal with a maximum of 1.915 V for the champion DAP + anneal tandem, 1.901 V for the champion control and 1.903 for the DAP. All the other solar-cell performance parameters maximize in the DAP + anneal with maximum FF of 0.70 (Control= 0.65 and DAP = 0.63) and $J_{SC}$ of 18.93 mA/cm² (Control= 18.86 mA/cm², DAP = 18.79 mA/cm²), resulting in PCE of 25.29% (Control = 23.26%, DAP = 22.72%).

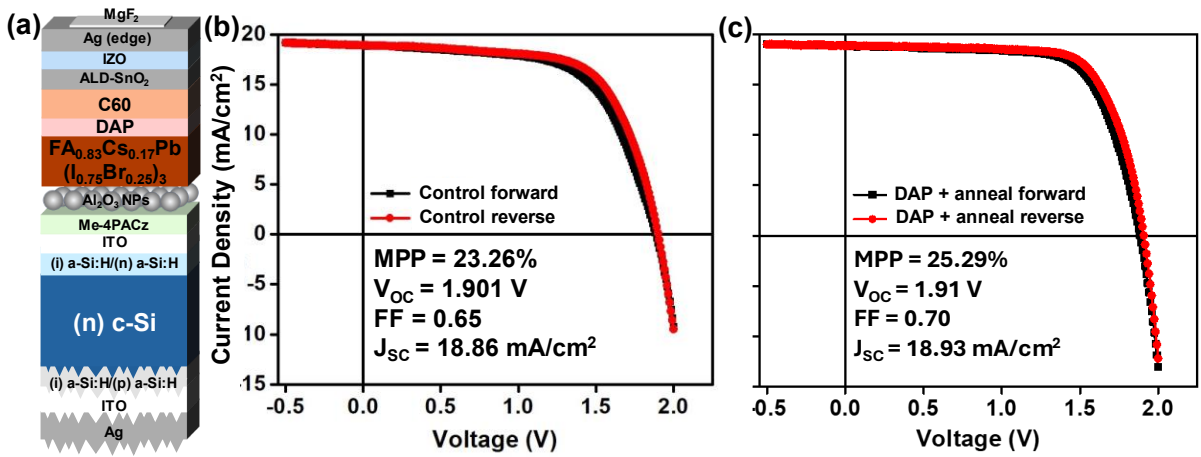

***Figure 4***: *a) Si-perovskite tandem architecture of devices made with the DAP surface passivation*

*step. b) Current-voltage (J-V) curve of the control Si-perovskite tandem and c) DAP + anneal. The values were recorded from 1 cm² tandem devices shown in the Inset of the figure.*

**Conclusion**

We demonstrate the efficacy of combining surface treatment with a diamine molecule and incorporating a post-processing step of annealing the $C_{60}$ interface to enhance the performance of 1.7 eV single-junction perovskite solar cells. Notably, we were able to implement this approach successfully into a silicon-perovskite tandem device improving the PCE from 23.26% to 25.29% and achieving a champion cell FF of 0.70 and $V_{OC}$ of 1.915 V. Moreover, we show that diamines react with formamidinium (FA) cations when deposited on the perovskite surfaces, consistent with our previous work on solution-based reactions of primary amines with FA cations. We note with interest the similarity between EDA, DAP, and the tail groups of the recently reported in recent amine-silane based passivation schemes and posit that similar reactions may take place.[48] We speculate these reactions may relieve defects caused by FA in interstitial spaces or anti sites at the surface. We attribute the improvement in $V_{OC}$ to the surface defect passivation effect of DAP and the improvement in FF to the post-annealing processing which enhances $C_{60}$ adhesion and promotes a more uniform perovskite composition throughout the film depth. This research underscores the ongoing efforts to overcome efficiency-limiting factors in silicon-perovskite tandem solar cells, offering insights into innovative processing strategies for future advancements. We anticipate the adoption of post processing annealing step in Si-perovskite tandem fabrication and its application to other interlayers to further enhance efficiency of tandems, while considering the possible chemistry of amine surface passivators with the formamidinium cation.

**Acknowledgments**

This work was primarily funded by the Office of Naval Research (Award # N00014-20-1-2587). ToF-SIMS was carried out at the Molecular Analysis Facility, a National Nanotechnology Coordinated Infrastructure site at the University of Washington which is supported in part by the National Science Foundation (awards NNCI-2025489, NNCI-1542101), the Molecular Engineering & Sciences Institute, and the Clean Energy Institute. The CEA team acknowledges the NEXUS project funded from the European Union's Horizon Europe research and innovation program under grant agreement No. 101075330 for the fabrication of the Si bottom cell used in

Si-perovskite tandem devices. Views and opinions expressed are, however, those of the author(s) only and do not necessarily reflect those of the European Union or RIA. Neither the European Union nor the granting authority can be held responsible for them. We thank the beamline scientists Honghu Zhang and Ruipeng Li at the 11-BM CMS beamline at NSLS-II, Brookhaven National Lab.